# Leveraging Social Media Sentiment for Predictive Algorithmic Trading Strategies

Gatik Goyal[1], Sharvil Phadke[2], Arnav Sharma[2], Huifang Qin[#]

[1]Irvington High School, Fremont, California, USA
[2]Cupertino High School, Cupertino, California, USA
[#]Advisor

ABSTRACT

This study investigates how social media sentiment derived from Reddit comments can be used to enhance investment decisions in a way that offers higher returns with lower risk. Using BERTweet we analyzed over 2 million Reddit comments from the subreddit r/wallstreetbets and developed a Sentiment Volume Change (SVC) metric combining sentiment and comment volume changes, which showed significantly improved correlation with next-day returns compared to sentiment alone. We then implemented two different investment strategies that relied solely on SVC to make decisions. Back testing these strategies over four years (2020-2023) our strategies significantly outperformed a comparable buy-and-hold (B&H) strategy in a bull market, achieving 70% higher returns in 2023 and 84.4% higher returns in 2021 while also mitigating losses by 4% in a declining market in 2022. Our results confirm that comment sentiment and volume data derived from Reddit can be effective in predicting short-term stock price movements and sentiment-powered strategies can offer superior risk-adjusted returns as compared to the market, implying that social media sentiment can potentially be a valuable investment tool.

## 1 Introduction

The financial market is a dynamic ecosystem influenced by numerous factors, including economic data, geopolitical events, and investor sentiment. Over the past decade, the rise of social media platforms such as Twitter and Reddit has introduced a new dimension to the analysis of market trends. Communities like r/wallstreetbets on Reddit have demonstrated the power of collective sentiment to drive significant price movements in stocks, underscoring the growing influence of retail investors. One such example was the GameStop short squeeze in 2021, where users from the subreddit r/wallstreetbets spurred a short squeeze on GameStop stock, skyrocketing its stock price by over 1,600% over the span of one month. This phenomenon presents an opportunity to leverage sentiment analysis on user-generated content as a predictive tool for stock investing.

This research aims to explore the relationship between sentiment expressed in Reddit comments and stock price movements. We decided to focus primarily on tech stocks in our research as these stocks commonly have much more discussion surrounding them on social media platforms such as Reddit. This study is centered around the hypothesis that there is a statistically significant correlation between market sentiment derived from social media and future stock trends that can be utilized to get higher returns than the market. By employing natural language processing (NLP) techniques, we extracted and analyzed the sentiment of millions of Reddit comments relevant to the stocks we include in our study. Using the extracted sentiment signals and daily comment volumes, we built multiple sentiment-informed stock investment algorithms and tested them on historical stock price data comparing the results to a conventional buy and hold strategy.

The findings of this research could have significant implications for the field of financial analysis and investing, particularly in understanding the role of crowd-sourced sentiment in shaping market dynamics. By comparing the returns of an investment algorithm relying on sentiment extracted from Reddit comments to a simple buy and hold strategy, this study demonstrates that social media sentiment can be an invaluable tool for stock price prediction and making informed stock investment decisions.

## 2 Literature Review

Sentiment analysis is a branch of natural language processing (NLP) that involves extracting subjective information from text such as emotions and opinions. Various techniques for sentiment analysis exist including lexicon-based methods which rely on predefined dictionaries correlating words with their sentiment, to machine learning based approaches that involve training models on labelled datasets (Kumar et al., 2023). In our study, the two specific approaches we utilized were Pysentimentio and ChatGPT-3. Pysentimentio is an open-source Python library that leverages transformer-based architectures like BERT and RoBERTa and is fine-tuned on sentiment-related tasks (Pérez et al., 2021). ChatGPT-3, on the other hand, is based on the GPT-3 architecture, a large-scale transformer model known for its contextual understanding and generative capabilities. Its large size allows it to handle nuanced or complex discussions and perform logical reasoning (Brown et al., 2020).

Sentiment analysis has found diverse applications, including analyzing customer feedback, monitoring political sentiment, and assessing public health trends. For instance, companies use it to gauge consumer satisfaction, while governments employ it to track public opinion on policies. The increasing use of social media platforms, such as Twitter and Reddit, has provided a rich data source for sentiment analysis, enabling real-time monitoring of public discourse (Soni Sweta, 2024). Recent trends include advancements in transformer-based models and the integration of sentiment analysis into broader decision-making pipelines. This evolution underscores its growing importance across industries (Wankhade et al., 2022).

The utility of sentiment analysis extends to financial markets, where online forums and social media offer insights into investor sentiment. In the past the possibility of whether the stock market could be predicted has been heavily debated. Early economists had agreed on the idea of the efficient market hypothesis and the "random walk" which was the belief that all available information about a stock was immediately priced in and thus all changes were completely random (Malkiel, 2003). However, more recent research has shown that markets do not actually follow the efficient market hypothesis and can be predicted to an extent (Butler & Malaikah, 1992). Researchers have investigated various approaches towards the goal of predicting price changes. Numerous studies such as those by Chan, Lee, and Liu have explored the possibility of utilizing market momentum to predict stock prices (CHAN et al., 1996; Lee & Swaminathan, 2000; Lui et al., 1999). These studies emphasize that the efficient market hypothesis is not entirely accurate, concluding that the market often "under reacts" to new information and that new information is not immediately priced in like the efficient market hypothesis claims. With recent advancements in natural language processing and sentiment analysis there have also been a rise in studies utilizing sentiment extracted from social media data to predict stock prices. These studies have had varying levels of success, studies including Darapaneni and Mokhtari found that sentiment-based predictions were not consistently accurate and did not have a strong correlation with actual stock prices.

While prior studies have explored the use of sentiment analysis in financial markets, they have produced mixed results, with some demonstrating the potential of sentiment signals for stock price prediction and others highlighting limitations in accuracy and consistency. These inconsistencies often stem from differences in data sources, sentiment extraction methods, and evaluation metrics. For instance, some studies have focused on specific events, such as the COVID-19 pandemic, or relied on limited sentiment sources like news articles or tweets, without providing a comprehensive evaluation of sentiment signals across diverse market conditions (Das et al., 2022; Nguyen & Dev, 2023). Additionally, many studies have not thoroughly examined the risk-adjusted performance of sentiment-based strategies or their robustness in both upward and downward market trends (Zhang & Skiena, 2025). Our research addresses these gaps by developing and testing a sentiment-powered investment strategy that leverages sentiment signals derived specifically from Reddit comments, a platform known for its active retail investor community. Unlike prior work, we provide a detailed evaluation of the strategy's performance, including both return and risk metrics, and assess its effectiveness across different market conditions. By comparing the results to a conventional buy-and-hold strategy, we demonstrate the potential of sentiment signals to consistently generate higher returns while maintaining similar levels of risk. Furthermore, our study offers insights into the quality of sentiment signals extracted from Reddit and highlights the best practices for integrating these signals into investment decision-making processes. These contributions provide a more nuanced understanding of the role of social media sentiment in financial markets and establish a foundation for future research in this area.

## 3 Data Retrieval and Processing

## 3.1 Downloading and Preprocessing social media data

In order to extract Reddit comments from the subreddit r/wallstreetbets, we downloaded a torrent that stored all comments to the subreddit since its creation on January 31, 2012. Using a Python script, we extracted all comments from the torrent that mentioned either the name or the NASDAQ symbol for ten pre-selected tech companies varying in volatility: Google, Tesla, Meta, Nvidia, Apple, eBay, Amazon, Netflix, Microsoft, Intel. We stored these extracted comments along with their timestamps and the companies they referenced in csv files.

## 3.2 Sentiment Analysis Techniques

### 3.2.1 Sentiment Extraction Models

We experimented with two different approaches to extracting the sentiment from these filtered Reddit comments: BERTweet and ChatGPT. BERTweet is an open-source sentiment model that was trained on a dataset of 800 million tweets and returns the probabilities of the given comment having positive, neutral, and negative sentiment (Pérez et al., 2021). Similar to BERTweet, ChatGPT is also an NLP model but with a vastly different architecture and purpose. BERTweet is a much smaller single purpose model that only performs text classification, while ChatGPT is a much larger autoregressive language model that can perform a wide range of tasks and responds in a conversational style. We decided to use these two models in particular, as we wanted to compare the effectiveness of a simpler, social media-trained sentiment model like BERTweet with a more complex AI model like ChatGPT.

### 3.2.2 Sentiment Analysis Implementation

Due to the differences between the two models, with ChatGPT being an LLM with a text-based output and BERTweet being a pure sentiment analysis model with a numerical output, the way we used them to extract sentiment was different as well. To extract sentiment using ChatGPT we utilized Open AIs API for gpt-3.5-turbo with the following prompt "You will be provided with a comment, and your job is to predict what the commenter thinks will happen to a stock price. Format your answer as a JSON with a stock's ticker symbol as the key and either 'Up', 'Neutral', or 'Down' as the value. Include each stock relevant to the comment provided" followed by a Reddit comment. This would result in ChatGPT outputting the stock tickers for any stocks mentioned in the comment followed by 'Up', 'Neutral', or 'Down'. These values were then converted into numerical values where 'Up' = 1, 'Neutral' = 0, and 'Down' = -1.

For BERTweet comments were provided to the model one at a time. The total sentiment was calculated by adding the positive score with half of the neutral score and then subtracting 0.5 to get a negative to positive range of -0.5 to 0.5. This sentiment score was considered the sentiment for all stocks mentioned in the comment which we noted when filtering the comments.

### 3.2.3 Sentiment Extraction Accuracy

In order to test and compare the two models, we picked 200 random comments from our dataset, which make up about 0.1% of all the dataset's comments, and manually labelled these as Positive, Negative, or Neutral. Then we compared the sentiment scores from the models with these human labels as the ground truth.

| **Sample Comments** | **BERTweet** | **Chat GPT** | **Human Labelling** | **BERTweet Match** | **GPT Match** |
|---|---|---|---|---|---|
| "MSFT the little engine that could. Keep climbing, I believe in you.  Boomers trying to kill it." | 0.4555097 | MSFT: "Up" | Positive | Yes | Yes |
| "NVDA may be in a bubble, and traders should exercise caution when investing." | 0.0086248 | NVDA: "Down" | Negative | No | Yes |
| "Just bought NVDA puts purely out of spite. Don't care if they're down." | -0.4029934 | NVDA: "Up" | Negative | Yes | No |

**Table 1: Sample of comments with sentiment scores.** Displays three sample Reddit comments from our dataset along with sentiment score extracted through BERTweet, the output from ChatGPT, and what we manually labelled the comment as. The last two columns indicate whether ChatGPT's and BERTweet's labels were accurate as compared to the manually assigned label.

### 3.2.4 Analysis and Conclusion

In our evaluation of sentiment analysis models, BERTweet achieved an 85% accuracy, slightly surpassing ChatGPT's 80% accuracy on our manually labeled comments. While both models demonstrated proficiency, a detailed comparison highlighted their distinct strengths and weaknesses. ChatGPT excelled in understanding the nuances of financial terminology and logical reasoning. For instance, it could accurately identify bearish sentiment in comments like "I am so happy I didn't invest in…" which BERTweet would misclassify as positive. Furthermore, the use of ChatGPT offered separate sentiment classifications when multiple stocks were referenced within a single comment, a capability that BERTweet lacked as it only outputs a single numerical value. Interestingly, ChatGPT also tended to label comments more positively, aligning well with human judgments for positive sentiments but less so for neutral or negative ones.

While BERTweet was somewhat limited in its sentiment analysis capabilities for complex comments, overall it still achieved a high accuracy when compared to human labelling, demonstrating that this wasn't a significant issue in practice. The biggest limitation from BERTweet was that in many cases it failed to recognize when comments were irrelevant to the companies mentioned, leading to unrelated comments being misclassified as strongly positive or negative. This indicated a lesser grasp of contextual understanding compared to ChatGPT. Despite ChatGPT's more sophisticated interpretive abilities in practice it was slightly less accurate than BERTweet overall.

Ultimately, as both approaches to sentiment analysis had similar accuracies when compared to human labelling, we opted to use BERTweet for the rest of our research due to its efficiency and cost-effectiveness. From our testing we found that it took about 90 seconds and cost about 2 cents to process 100 comments from our dataset using ChatGPT. Thus, processing our extensive dataset of over 2 million comments with ChatGPT through OpenAI's API would have been significantly more difficult, costing over $400 and requiring approximately 500 hours to complete. Given BERTweet's slightly higher accuracy and its vastly more efficient processing capabilities for large volumes of data, it was clearly better suited for our study.

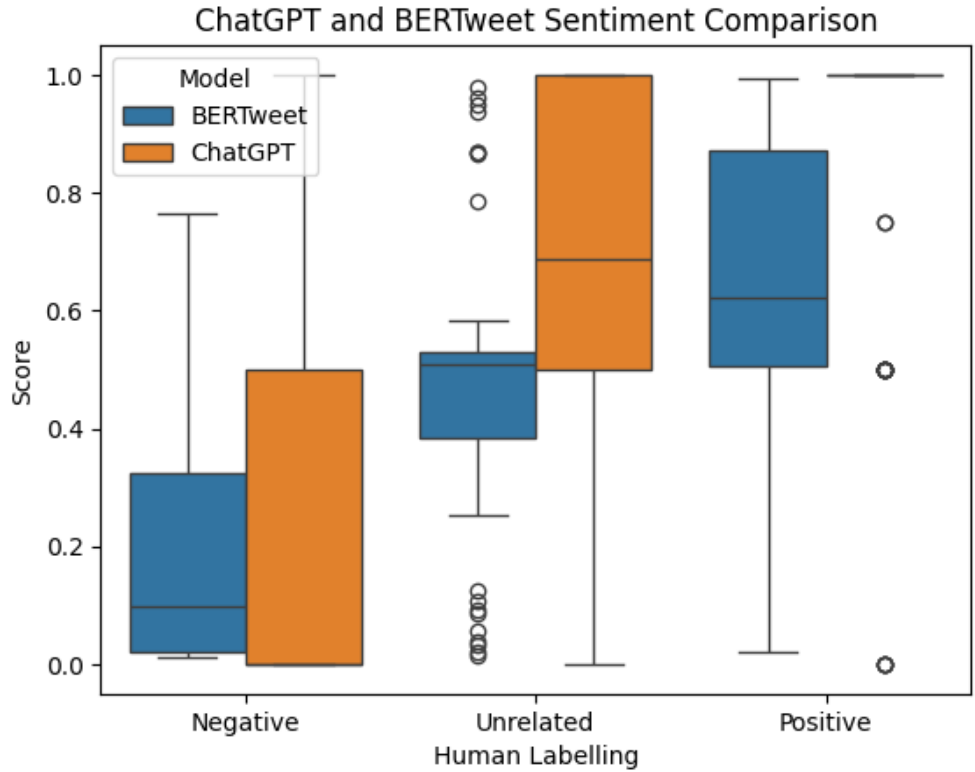

**Figure 1: ChatGPT and BERTweet labelling distribution:** The box and whisker's plot shows the values that ChatGPT and BERTweet output for the comments that were manually labelled as negative, unrelated, and positive. Here the ChatGPT outputs were converted to a numerical measurement so they could be easily compared with the output from BERTweet; "Up" became 1, "Neutral" became 0.5, and "Down" became 0. We also added 0.5 to BERTweet to get a similar range from 0 to 1.
Note: ChatGPT's sentiment scores for positive comments were consistently 1.0 (first quartile, median, and third quartile), showing a strong alignment with human judgment for positive comments.

### 3.3 Testing Predictive Power of Sentiment Signals

#### 3.3.1 Metric 1: Measuring Sentiment Change

With the sentiments of all the comments stored, we began to test the market potential of the comment sentiments by measuring the correlation between sentiment change and proceeding stock change. To do this, we went through every company for each year, measuring the change in the average sentiment values of comments mentioning the company from one day to one day after. We then compared this to the percentage change in that company's stock from one day after to two days after.

We displayed the results in a regression plot and calculated the correlations R value, p-value, and slope, to which we found minimal evidence of daily sentiment change as a viable stock predictor. The p-value ($0.036 < 0.05$) showed that the correlation was statistically significant, and not just based on chance; however, as per the low $R^2$ value (0.0006), the change in sentiment alone could barely explain the variation in stock growth.

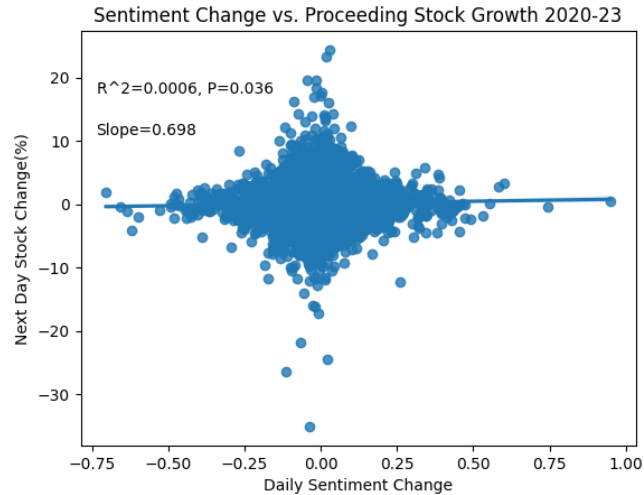

**Figure 2: Change in Sentiment versus Next Day Stock Growth:** Shows the relationship between the daily change in sentiment and the proceeding change in the stock price for every day from 2020-2023. Each point represents the sentiment and price change from one day to the next for a single company, with the x-axis showing the percentage change in sentiment and the y-axis showing the percentage change in stock price.

### 3.3.2 Metric 2: Measuring Sentiment Volume Change

Since the daily sentiment change had a very low correlation with the proceeding stock growth, we instead opted to create a more meaningful measurement we coined the "Sentiment Volume Change" (SVC). That is, instead of taking the change in daily mean sentiment, we took that change in the mean sentiment from day-to-day and multiplied it by the absolute value of the change in the number of comments. The purpose of the SVC is to measure sentiment change while accounting for the change in the volume of comments which could indicate increased volatility or new information regarding the company becoming available to the public. A change in how much investors are discussing a stock can be considered a good reflection on future stock growth if the investors are talking about the stock in a more positive way, and a bad reflection if they are talking about the stock in a more negative way. The SVC reflects this idea; it is positive if the mean sentiment has increased and negative if it has decreased, and a higher change in the number of Reddit comments mentioning a company increases the measurement's magnitude.

Using the same method of correlation measurement as before, we applied the SVC and compared it to next day stock growth, and found a statistically significant correlation (p-value < 0.05) with a higher $R^2$ value (0.0214), indicating that the SVC could explain more variation in stock growth than sentiment alone. What particularly caught our attention was that the correlation seemed to increase further from the origin. To test this theory, we removed comments that had an SVC within 20 units from the origin, and got a statistically significant correlation (p-value < 0.05) which could explain 13.04% of the variation of stock growth ($R^2 = 0.1304$). This was the foundation for our first investment strategy.

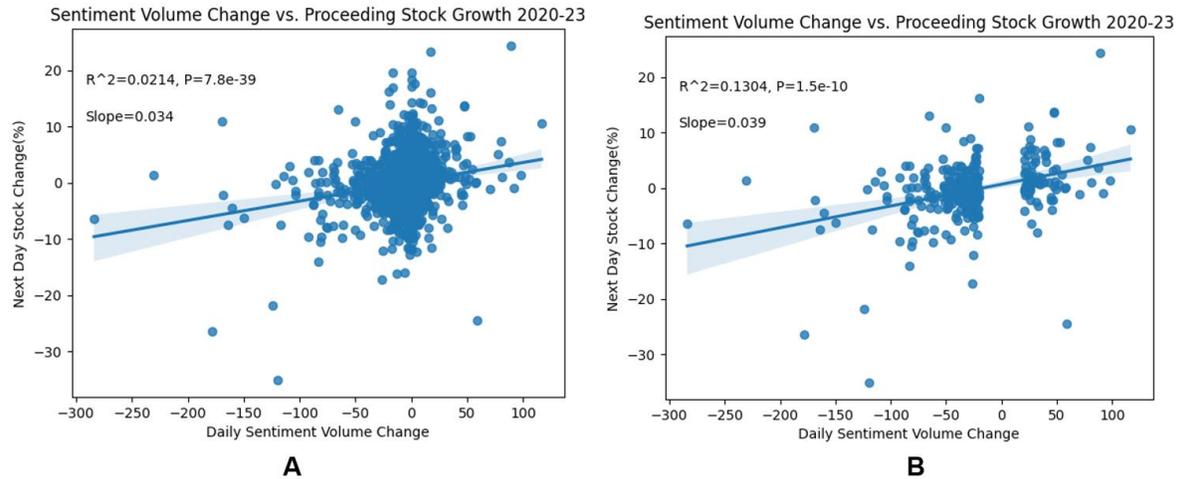

**Figure 3: Sentiment Volume Change versus Next Day Stock Growth:**
Figure 3a) shows the regression plot that contains one data point for each company and each day from 2020-2023. This time we have the Sentiment Volume Change metric for each point on the x-axis; the y-axis is the same. Figure 3b) shows the Sentiment Volume Change versus Next Day Stock Growth (excluding SVC values within (-20,20)). From this plot we can see that the more extreme SVC values have a higher correlation with stock price changes.

# 4 Single Stock Investment Algorithm

4.1 Investment Algorithm Implementation

*4.1.1 Baseline Sentiment Strategy*

For the first investment strategy, we supply $100 for an individual stock. $50 of this is put in a savings "account", and the other $50 is invested directly into the stock. Based on our previous analysis, extreme SVC values have a statistically significant correlation with future stock growth or decline. As such, the core of this strategy is transferring money between the investment and the savings if the SVC gives a strong indication of stock growth or decline. Every day, for an entire year, after the stock market closes, the SVC is calculated based on the stored Reddit comment data about the stock. If the SVC is greater than a predetermined positive threshold, indicating future stock growth, then all of the savings are transferred to the investment account. If the SVC is less than a predetermined negative threshold, indicating future stock decline, then all of the investments are transferred to the savings. The following day, the transaction is reversed. If it is not possible to reverse the exact monetary transaction, then all available money is transferred back to the account that previously lost money. This way, in the event that the market crashes and there's no money left in the stock, the algorithm will not crash; instead, no money will be transferred.

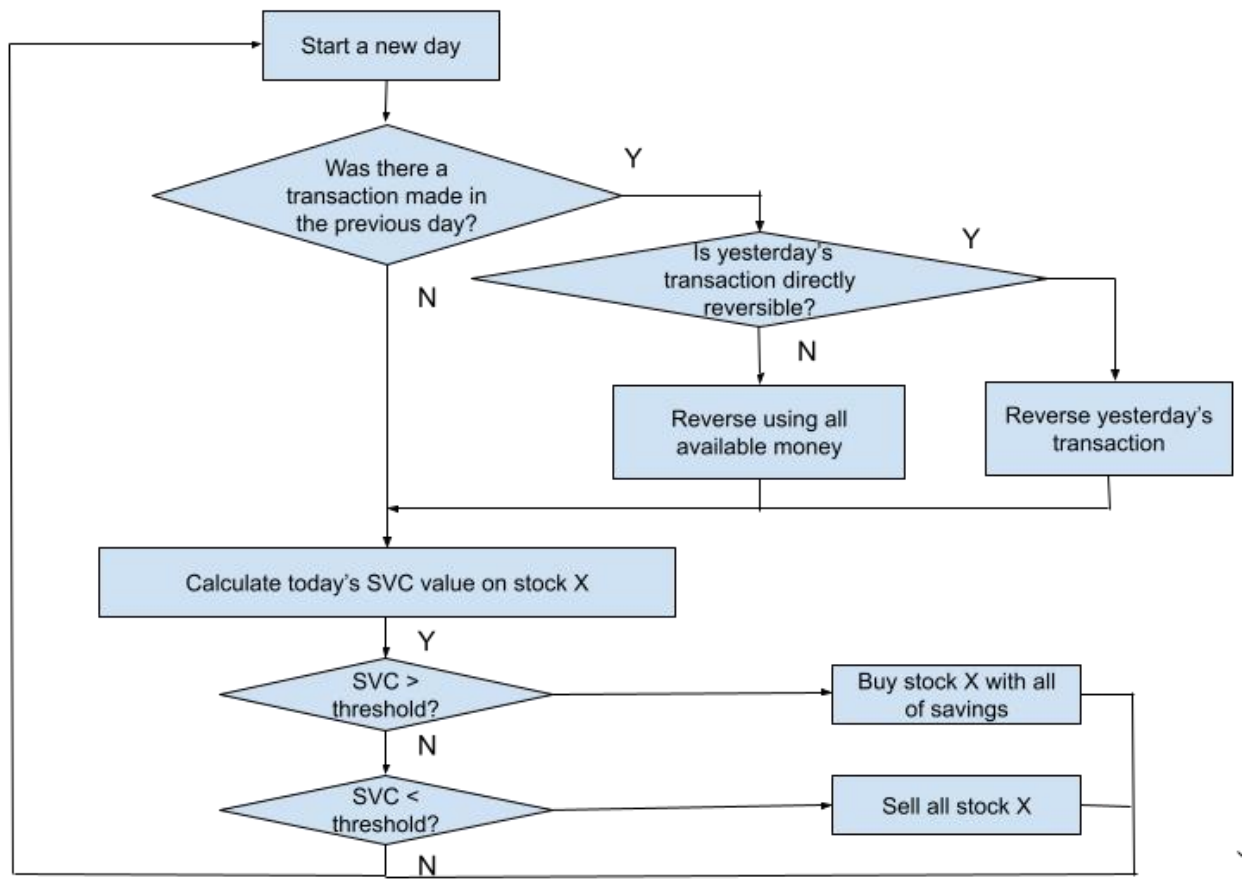

**Figure 4: Sentiment Strategy Visualization:** Block diagram illustrating the daily sentiment strategy for a hypothetical stock X.

### 4.1.2 Baseline Buy and Hold Strategies

In order to provide a base comparison for the sentiment-based algorithm's performance, we developed the $50 and $100 "Buy and Hold" algorithms, which are both also given $100 initially. The $50 B&H stores $50 in each stock's savings and $50 in each stock but never makes any changes from there. The $100 B&H stores all $100 in each stock but never makes any changes from there. By comparing the sentiment algorithm to B&H ones, we wanted to get a relative assessment of the algorithm's effectiveness in balancing risk and return.

### 4.1.3 Sentiment Thresholds

Sentiment thresholds were chosen by testing a large set of potential threshold pairs on the strategy during February 1 - April 1, 2020, when the stock market significantly declined, and April. 1 2020 - July 1. 2020, when the stock market significantly grew. We applied each threshold pair to the algorithm in both time periods and measured its gain from the $50 B&H strategy. The results for each time period are displayed in the diagram below. The gains in April - July 2020 were higher than the gains in February - April 2020, so in order to standardize the metric, we converted each pair's monetary gains to percentiles within each time period. Then, we calculated each pair's combined percentile from the two time periods and chose the pair with the highest result. With these restrictions, we attained a sentiment threshold pair of (10, -15).

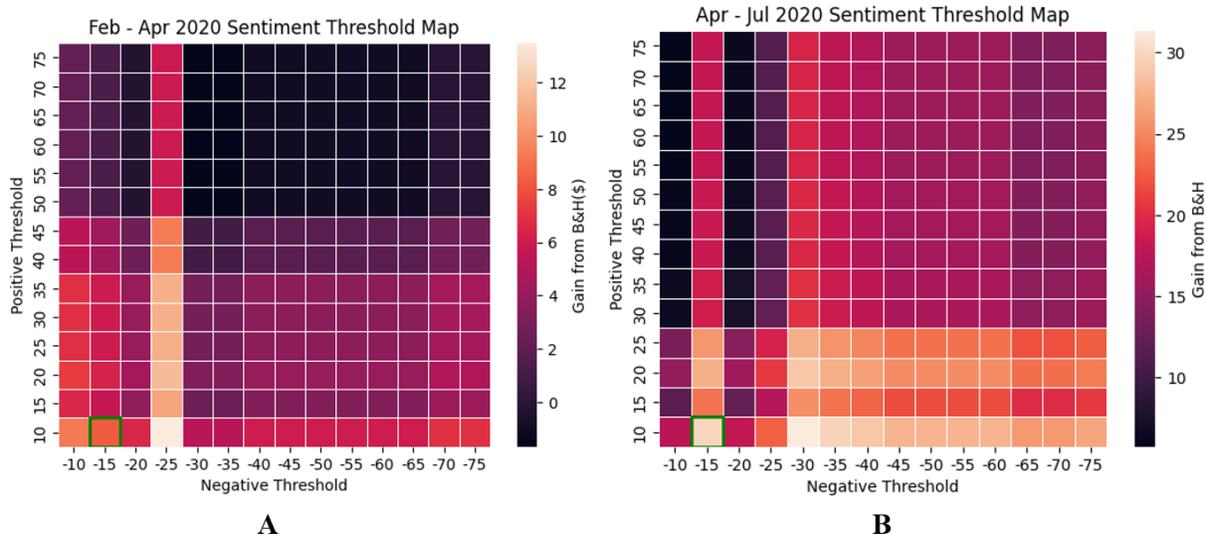

**Figure 5: Sentiment Threshold Results Heatmap:** The heat maps show the monetary gain of the sentiment-based algorithm when compared to the $50 B&H strategy, based on different threshold pairs. The left heatmap is applied to February 1, 2020 - April 1, 2020, which is when the stock market was falling. The right heatmap is applied to April 1, 2020 - July 1, 2020, which is when the stock market was growing. The green borders show the chosen threshold pair based on the metric explained above.

## 4.2 Strategy Results

### 4.2.1 Success Metrics

To assess the results of the sentiment algorithm, we developed a risk metric and a return metric.

The risk metric was measured as the standard deviation of an algorithm's percentage growth in total money after each day, with a higher standard deviation indicating larger changes in growth. We decided to use standard deviation for risk because it showed similar results to the other risk metrics we attempted – percentiles and confidence intervals – while also being more easily interpretable. It shows how the returns deviate from the mean return and thus how consistent a trading strategy is.

The return metric was calculated by subtracting the algorithm's initial total money from its total at the end of the year.

### 4.2.2 Algorithm Results

To assess the algorithm, we compared its performance from 2021-2023 to the $50 and $100 B&H algorithms. We decided not to apply the algorithm to 2020, as 2020 stock data was used to pick the sentiment thresholds.

The results showed that while the $100 B&H algorithm thrived comparatively in 2021 and 2023, when the stock market was in an upturn, it performed worse than the other two algorithms in 2022, when the market was in a downturn. The algorithm also had a consistently higher risk measurement in all three years, indicating that it is a high-risk algorithm that performs well only in good market conditions. Meanwhile, while the $50 B&H algorithm performed worse than the $100 B&H in 2021 and 2023, it lost less in 2022, while also having a lower risk for all three years. In comparison, the sentiment algorithm had a low-risk measurement nearly identical to the $50 B&H's risk but also performed slightly better than the $50 B&H in all three years, and was the best out of all three algorithms in 2022. For this reason, we concluded that the sentiment-based strategy was comparatively low-risk and high reward.

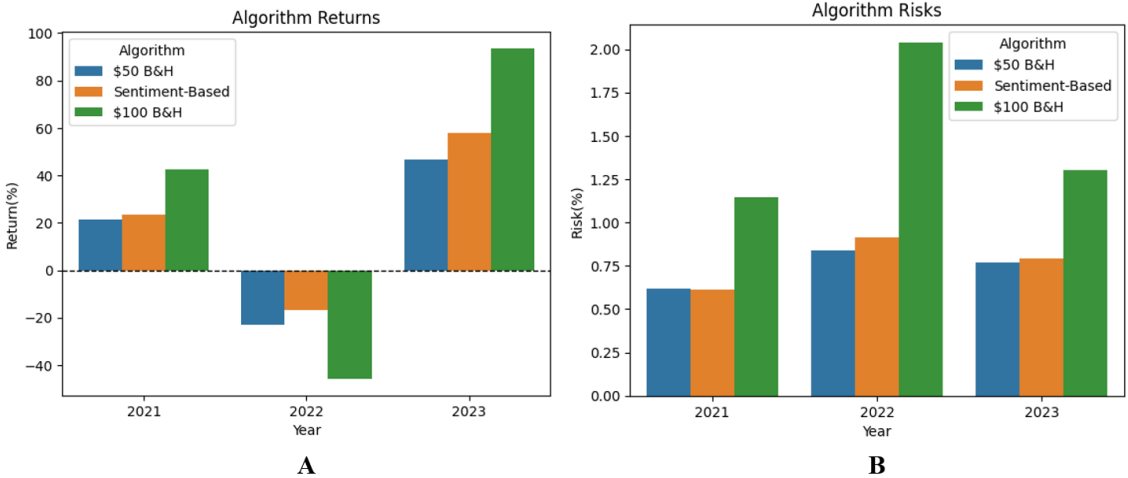

**Figure 6: Algorithm Performance Results**
Figure 6a) shows a bar graph of the returns, in percentage, for each of the algorithms from 2021-2023. Figure 6b) shows a bar graph of the risks, in percentage, for each of the algorithms from 2021-2023.

|  | $50 B&H Alg | Sentiment-Based Alg | $100 B&H Alg |
|---|---|---|---|
| **2021** | 21.3% | 23.4% | 42.6% |
| **2022** | -22.9% | -16.7% | -45.8% |
| **2023** | 46.8% | 58.0% | 93.5% |

**Table 2: Algorithm Returns:** The percentage return of each of the algorithms from 2021-2023, displayed as a numerical table.

|  | $50 B&H Alg | Sentiment-Based Alg | $100 B&H Alg |
|---|---|---|---|
| **2021** | 0.62% | 0.61% | 1.14% |
| **2022** | 0.84% | 0.92% | 2.04% |
| **2023** | 0.77% | 0.79% | 1.30% |

**Table 3: Algorithm Risk:** The percentage risk of each of the algorithms from 2021-2023, displayed as a numerical table.

### 4.2.3 Sentiment Action Analysis

An example of money transfers occurring due to a change in SVC is displayed below, where the specific stock was Microsoft during 2021. For a majority of days, the SVC remained relatively constant, and no further action was taken. However, 294 days after the start of investment, the SVC fell above the positive sentiment threshold, which led to the algorithm transferring all the storage's money to the investment temporarily. This led to the 4.21% proceeding stock growth to be fully capitalized on.

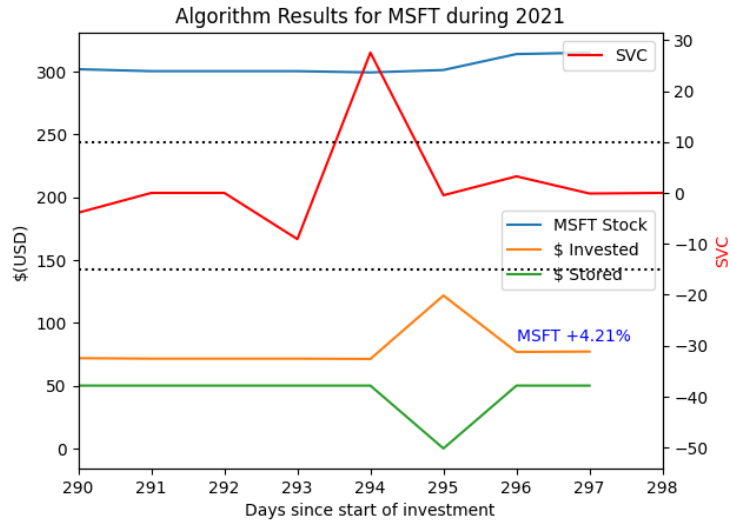

**Figure 7: Sentiment Analysis Graph for Microsoft:** Graphical representation of the money transfer explained above for Microsoft's 2021 stock, going from 290 to 297 days since the start of the investment. The stock value is shown in blue, the SVC value in red, the investment account in orange, and the storage account in green. The dotted horizontal lines represent the negative and positive sentiment thresholds. The blue text displays the stock growth the day after the transfer took place.

*4.2.4 Effectiveness of the Algorithm*

The consistent performance of the sentiment algorithm when compared to the $50 B&H indicates that there is potential for Reddit sentiment to be utilized to make more informed investment decisions. It also shows that online communities pertaining to the stock market, like r/wallstreetbets, do reflect to a partial degree the general future sentiment about stocks among all investors.

However, there are a few clear areas of improvement for this preliminary algorithm. For one, the algorithm does not fully utilize the $100 initially given, which leads to it significantly underperforming the $100 B&H in 2021 and 2023, which were years where the stock market rose. Additionally, the algorithm only works on one stock at a time. This means that when the SVC indicates that a stock is going to rise or drop significantly, the algorithm cannot utilize the money from other stocks. We took these areas of improvement into account as we developed our second iteration of the sentimental algorithm, shown in the next section.

# 5 Multi-stock investment algorithm

In our second iteration of the investment strategy, we opted to take an approach that redistributes capital among multiple stocks every day, so that we could evaluate sentiment's effect when combined with diversification. We also aimed to create an investment algorithm that always uses sentiment instead of using it conditionally, unlike our previous algorithm.

5.1 Investment Strategy

In our algorithm design, we prioritized simplicity and direct use of sentiment. We continued to use the SVC metric as it captured both the strength and volume of the sentiment signal as equally important. The algorithm normalizes the SVC so that the sum of the outputs of the algorithm for each company is 100%, as shown by the table.

| **Steps** | **Stock I** | **Stock II** | **Stock III** | **Comment** |
|---|---|---|---|---|
| **1. Measure SVC** | 7.3 | 9.1 | -9.2 | Taking SVC as INPUT |
| **2. Scale SVC to [0, infinity) range** | 16.5 | 18.3 | 0 | Bounds our signal so that we don't try to invest a negative amount |
| **3. Scale SVC to [0,100%) range** | 47.4% | 52.6% | 0% | OUTPUT to be used in investment strategy |

**Table 4: Multi-Stock Investment Strategy Example**
The table shows a sample calculation for the distribution of capital for one day given the previous day's SVC change.

The algorithm finds the minimum SVC value and then adds that minimum to all the values. Then, it divides by the sum of the SVC value to produce a list of percentages that add up to 100%. By normalizing our theoretically unbounded SVC value to a range that adds up to 100%, we can invest the output percentage correlated with each stock. If there is no data for a stock, it will be treated as having 0 sentiment change. In the edge case that there is no data for all the stocks or that the sentiment volume change for all the stocks is the same number, no reallocation of capital occurs.

## 5.2 Results

### 5.2.1 Comparison to Sentiment Baseline Algorithm

To obtain an overall view of our success before more deeply analyzing specific aspects of our algorithm's success, we compared our strategy to the common strategy, "100% Buy and Hold" which initially splits the money equally between the 10 stocks and does not make any trades after that. Analyzing over the past 4 years shows that our multi-stock sentiment powered investing algorithm averaged 89.324% higher return per year, with 43.978% higher statistical risk.

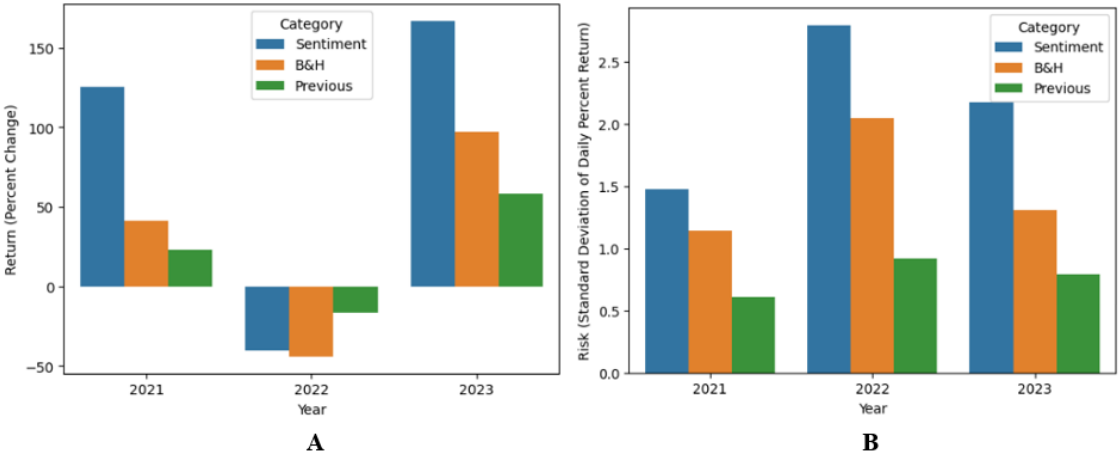

**Figure 8: Multi-stock Algorithm Performance Results**
Figure 8a shows a bar graph of the returns, in percentage, for the Multi-stock (Blue), Buy and hold (Orange) and single stock (Green) algorithms for each year from 2021 to 2023. Figure 8b shows a bar graph of the risks, in percentage, for each of the algorithms from 2021-2023.

|      | Multi-stock Alg. | Single-stock Alg. | $100 B&H |
|------|------------------|-------------------|----------|
| 2021 | 125.6%           | 23.4%             | 41.2%    |
| 2022 | -40.1%           | -16.7%            | -44.1%   |
| 2023 | 166.9%           | 58.0%             | 96.9%    |

**Table 5: Algorithm Returns**
The percentage return of each of the algorithms from 2021-2023, displayed as a numerical table.

|      | Multi-stock Alg. | Single-stock Alg. | $100 B&H |
|------|------------------|-------------------|----------|
| 2021 | 1.47%            | 0.61%             | 1.14%    |
| 2022 | 2.79%            | 0.92%             | 2.04%    |
| 2023 | 2.17%            | 0.79%             | 1.30%    |

**Table 6: Algorithm Risk**
The percentage risk of each of the algorithms from 2021-2023, displayed as a numerical table.

### 5.2.2 Analysis through Permutation of Stocks

In order to compare how the number of stocks being considered affects the results of the strategy we decided to test the data on all combinations of different numbers of stocks, going from 1 stock to 10 stocks. For example, when testing the performance of the algorithm for k stocks we would run the algorithm on 10 choose k stocks and then take the average of the returns for each test. The number of stocks is important as we wanted to gauge how sentiment behaved while diversified, requiring us to analyze the algorithm over a multitude of different starting sets of stocks. The graphs below show the mean return, risk and return to risk ratio among all the combinations that contain a given number of stocks for both buy and hold and the return.

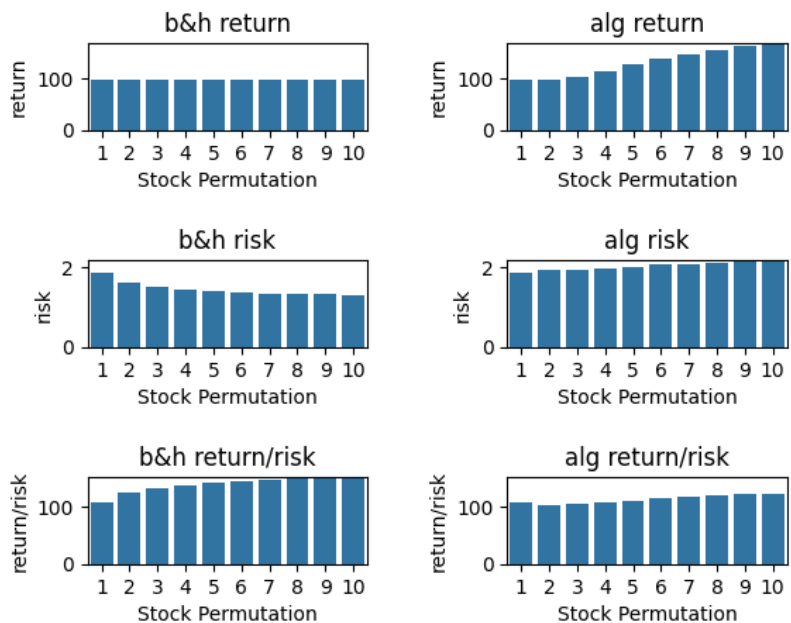

**Figure 9: Permutation analysis graph set for 2023**
Figure 9 shows the results of the permutation experiment for 2023. The subplots to the left, labeled "b&h", show the results for Buy and Hold. The results to the right, labeled "alg" show the multi-stock algorithm results. From top to bottom the rows show the return, risk, and return to risk ratio for each algorithm.

The results for Buy and Hold are consistent with known information about the strategy: diversification does not lower the expected return, but does reduce the risk, with diminishing effectiveness.

Comparatively, the sentiment algorithm's return increases with diversification, while risk also increases. As shown by our sentiment score analysis, the sentiment volume change has predictive power for the stock it's associated with, so having more stocks increases the total opportunities for a sentiment-based algorithm to catch an upwards trend in a particular stock. The overall return over risk is lower for 2+ stocks, meaning that if someone equally valued statistically low risk and high return, buy and hold would outperform the multi-stock algorithm. The higher statistical risk is likely due to the algorithm highly favoring stocks with high SVC on a particular day due to the algorithm design; the SVC itself is highly volatile, meaning that the algorithm very commonly shifts nearly all of its invested capital between stocks, leading to higher volatility. The volatility increase can be explained by the lower chance for the investment algorithm to keep capital invested the more stocks there are, as it becomes increasingly likely that a different stock than the current highest investment will have the highest SVC the next day than the current stock.

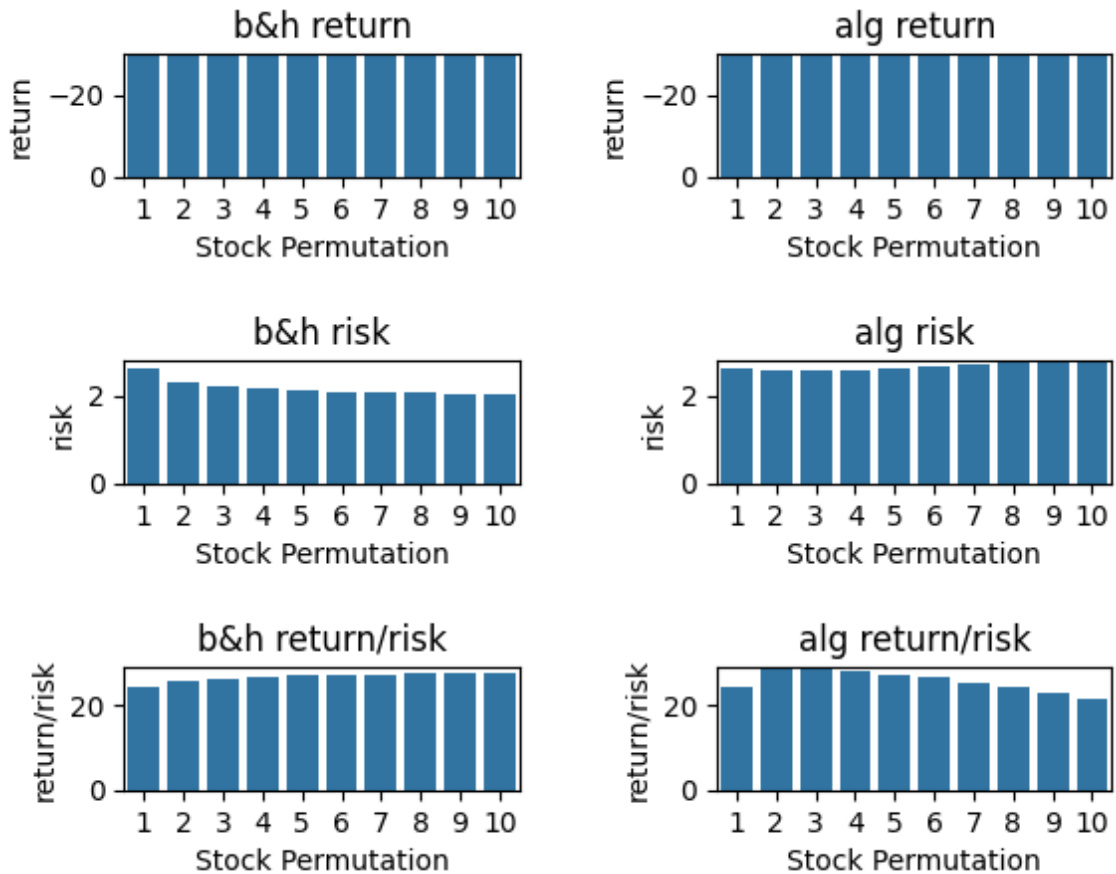

**Figure 10: Permutation analysis graph set for 2022**
Figure 10 shows the results of the permutation experiment for 2022. The subplots to the left, labeled "b&h", show the results for Buy and Hold. The results to the right, labeled "alg" show the multi-stock algorithm results. From top to bottom the rows show the return, risk, and return to risk ratio for each algorithm.

During 2022, the stock market showed extreme downwards trends especially for tech stocks, so it represents an extreme downwards market. In this investment situation, we see very different trends. The algorithm return is virtually constant, unlike in 2023. The risk shows the same upward trend as 2023. These graphs show that our multi-stock algorithm is significantly hindered by the lack of an upwards market, showing that sentiment is most effective when used in an upwards market.

### 5.2.3 Logging and Validation

By producing a distribution of the percent gain or loss per day, we further understand risk and volatility more than by just the standard deviation of percent increase. The following graphs show the distribution of percent increases for 2023.

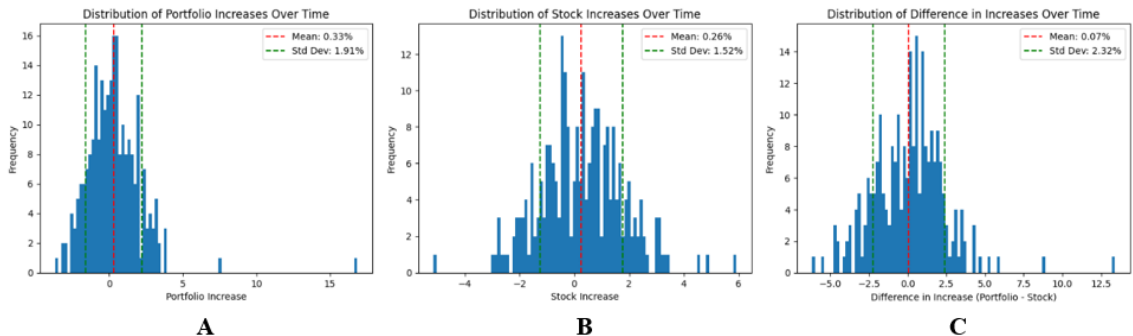

A  B  C

**Figure 11: Distribution of portfolio Increases over 2023**
Figure 11 shows the distribution of daily percent increase of the total capital in the investment portfolio each day. Figure 11a to the left shows the results for the multi-stock algorithm, figure 11b in the center shows Buy and Hold, and figure 11c shows the differences between the two algorithms. The means and standard deviations are shown by green and red lines respectively.

These graphs demonstrate how the sentiment algorithm uses the positive prediction power of the sentiment scores – shown in Figure 2 and Figure 3 – to achieve a higher return than buy and hold during 2023. Our sentiment scores were observed to be more accurate at the extremes, which is reflected in the graph as the upper outliers are much higher than in buy and hold. This is because the algorithm was able to correctly predict and capitalize on large increases in stock price that the sentiment data was able to predict. By looking at the difference between the multi-stock sentiment algorithm and the buy and hold returns per day, we can confirm that These increases are due to logical investments rather than higher by chance: the mean of the different is positive showing that on average our sentiment strategy yielded 0.07% higher return per business day, and the upper extremes are far more deviant from the mean than the lower extremes.

A potential bias our team considered was that the multi-stock algorithm was picking out stocks that were doing well, and they happened to continue to do so. One way we alleviated this in our algorithm design was using the SVC signal rather than just sentiment scores or comments. To validate that this approach did remove this form of bias, we compared the average percent investment in each of the stocks, and compared it to the increase during the year for the included stocks so we could identify potential bias.

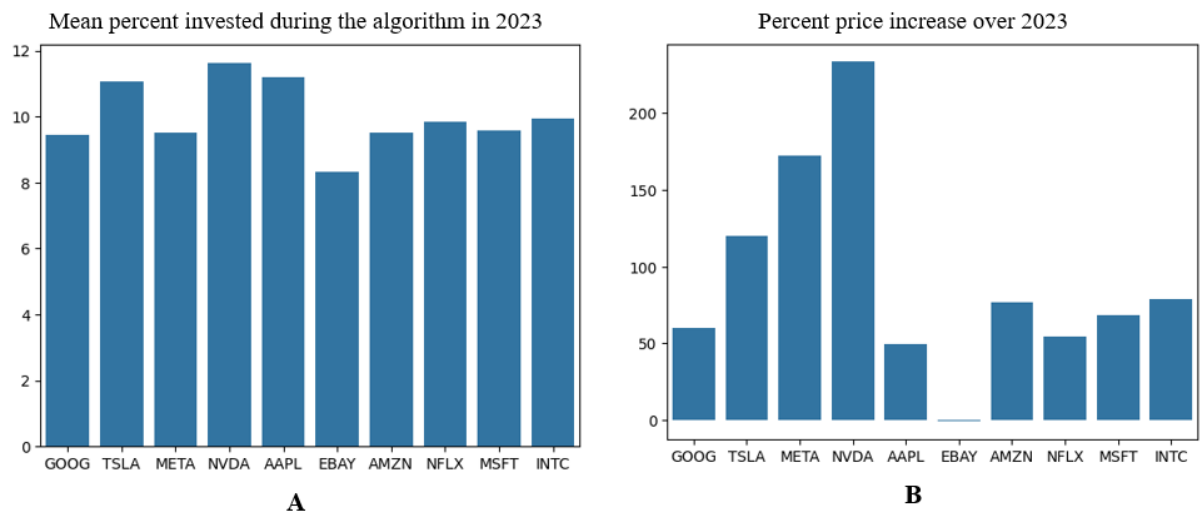

**Figure 12: Total investment analysis**
Figure 12a (to the left) shows the mean percent invested during 2023, adding up to 100%, calculated by averaging the percentage invested in each stock each day. Figure 12b shows the price increase over 2023 for each stock.

While there are extremes, the standard deviation of the mean percent invested values is 1.140, showing that the algorithm was fairly equitable in using all the stocks provided. When investing using the Buy and Hold strategy with the same distribution, the return is only 4.171% better, showing that our algorithm was not excessively biased towards well performing stocks and that the increased return was not due to stock selection bias.

## 6 Conclusion

This study investigated the predictive power of sentiment analysis on Reddit comments for stock price movements and its potential to inform investment strategies. By utilizing natural language processing techniques, specifically the BERTweet model, we quantified the sentiment of user discussions on r/wallstreetbets and developed an investment strategy leveraging sentiment-driven signals. Our analysis found that while daily sentiment changes alone did not significantly correlate with stock price movements, a metric involving both sentiment and comment volume showed higher correlations especially for extreme values.
Through back testing, we demonstrated that a sentiment-informed trading strategy based on SVC outperformed a traditional buy-and-hold approach, generating higher returns over the tested period. These results suggest that social media sentiment, when analyzed properly, can serve as a valuable complementary tool for investment decision-making. Future research could expand on this work by incorporating additional data sources, such as Twitter or financial news sentiment, improving model architecture, and testing sentiment-based strategies across different market conditions. Further exploration into reinforcement learning and deep learning models for trading decisions could enhance the robustness of sentiment-driven investment strategies. As financial markets continue to be shaped by retail investor sentiment and online discussions, sentiment analysis will likely play an increasingly important role in modern quantitative finance.

## 7 Limitations

One of the main limitations of this study is its reliance on Reddit comments from r/wallstreetbets. While this subreddit offers a rich source of retail investor sentiment, it may not be representative of the broader market sentiment or other influential investor groups. The specific demographic and trading habits of r/wallstreetbets users could introduce a bias, limiting the generalizability of the findings to the wider stock market. Furthermore, when filtering comments for our dataset, we did not consider the context of what a comment was replying to, potentially leading to misinterpretations of sentiment where a comment might be sarcastic or responding to a negative statement in a positive light. The focus on tech stocks (Google, Tesla, Meta, Nvidia, Apple, eBay, Amazon, Netflix, Microsoft, Intel) may also restrict the applicability of sentiment-driven strategies to other sectors, as different industries might exhibit varying sensitivities to social media sentiment. Lastly, the analysis is limited to data from the past four years

(2020-2023), which, while encompassing diverse market conditions, might not fully capture long-term trends or the efficacy of the strategies across even longer historical periods. Future research could explore diverse social media platforms, incorporate conversational context, broaden the range of stock sectors, and extend the historical data analysis to enhance the robustness and applicability of these sentiment-based trading strategies.

# Acknowledgements